\documentclass[journal]{vgtc}                     

\onlineid{0}

\vgtccategory{Research}

\vgtcpapertype{please specify}

\title{Visualising category recoding and numeric redistributions}

\author{\authororcid{Cynthia A.~Huang}{0000-0002-9218-987X}}

\authorfooter{
	  \item
  	Cynthia A. Huang is with Department of Econometrics and Business
Statistics, Monash University, Melbourne, Victoria, Australia.
  	E-mail: cynthia.huang@monash.edu
	}

\abstract{This paper proposes graphical representations of data and
rationale provenance in workflows that convert both category labels and
associated numeric data between distinct but semantically related
taxonomies. We motivate the graphical representations with a new task
abstraction, the \emph{cross-taxonomy transformation}, and associated
graph-based information structure, the \emph{crossmap}. The task
abstraction supports the separation of category recoding and numeric
redistribution decisions from the specifics of data manipulation in
ex-post data harmonisation. The crossmap structure is illustrated using
an example conversion of numeric statistics from a country-specific
taxonomy to an international classification standard. We discuss the
opportunities and challenges of using visualisation to audit and
communicate cross-taxonomy transformations and present candidate
graphical representations.}


\graphicspath{{figs/}{figures/}{pictures/}{images/}{./}} 

\usepackage{mathptmx}                  
\usepackage[numbers]{natbib}
\usepackage{fancyvrb}
\providecommand{\tightlist}{%
  \setlength{\itemsep}{0pt}\setlength{\parskip}{0pt}}
\usepackage{graphicx,grffile}
\makeatletter
\def\maxwidth{\ifdim\Gin@nat@width>\linewidth\linewidth\else\Gin@nat@width\fi}
\def\maxheight{\ifdim\Gin@nat@height>\textheight\textheight\else\Gin@nat@height\fi}
\makeatother
\setkeys{Gin}{width=\maxwidth,height=\maxheight,keepaspectratio}

\begin{document}
\maketitle

\hypertarget{introduction}{%
\section{Introduction}\label{introduction}}

Data are often collected and organised in classification structures.
From taxonomic plants and animals, to classification codes for product
types and occupations, defining and applying taxonomies is a standard
part of collecting data from the world around us. Unfortunately, without
some universal and timeless taxonomy of the world, there are inevitably
discrepancies in how identical or related phenomena are captured as
data. This leads to various opportunities for harmonising and
integrating semantically related but structurally distinct data,
i.e.~multi-schematic heterogeneous datasets
\citep{salonenChallengesHeterogeneousWeb2013}.

Ex-post harmonisation involves transforming data collected under
different designs or schemes and integrating them into a single dataset
for joint analysis \citep{kolczynskaCombiningMultipleSurvey2022}. This
process is commonly required in the social sciences to combine scarce
data into analysis-ready datasets. Ex-post harmonisation and associated
\emph{cross-taxonomy transformations} give rise to a variety of
provenance information. In this work, we focus on capturing and
visualising decisions made when recoding and redistributing category
indexed numeric variables between related taxonomies. To this end, we
briefly introduce a graph-based information structure for capturing
category recoding and numeric redistribution decisions, the
\emph{crossmap}, before discussing visualisation challenges and
opportunities.

Crossmaps are based on the observation that relationships between
category labels and numeric data from \(k\) taxonomic levels can be
represented as a \(k\)-partite graph. A taxonomic level, hereafter
referred to as the \emph{taxonomy} for brevity, can come from the same
hierarchical taxonomy or related taxonomies. These relationships can be
visualised as multi-layer graphs, with each taxonomy occupying a single
layer. As such, for a given transformation step, any standard bi-partite
graph visualisation can communicate which categories are connected to
each other, and weights on each link can capture the transformation of
numeric value between the two taxonomies. However, targeted use of
additional visual channels and strategic node ordering can also convey
additional provenance information. Visualisation can support
communication and discussion of core assumptions used to design a
particular mapping, as well as decisions which most impact downstream
analysis.

We proceed by reviewing some related areas of research, and providing an
example cross-taxonomy transformation with one-to-one and one-to-many
relations to demonstrate the crossmap format. We then examine the role
of visualisation in recording and communicating decisions and domain
expertise used to integrate data with related but distinct taxonomic
indices. Leveraging these observations, we propose and discuss graphical
representations of crossmaps.

The main contributions of this paper are:

\begin{itemize}
\tightlist
\item
  providing a provenance task abstraction for categorical recoding and
  numeric redistribution transformations in the context of ex-post data
  harmonisation, i.e.~the \emph{cross-taxonomy transformation};
\item
  outlining an associated graph structure for encoding mappings between
  taxonomies, i.e.~the \emph{crossmap};
\item
  examining the role of visualisation in recording and communicating
  decisions and domain expertise used to integrate data with related but
  distinct categories;
\item
  proposing graphical representations of crossmaps.
\end{itemize}

\hypertarget{background}{%
\section{Background}\label{background}}

Following Munzer's nested model for describing visualisation design
contributions \citep{munznerNestedModelVisualization2009}, this work can
be linked to a number of existing areas of data provenance and
visualisation research. At the domain problem and operation abstraction
levels, it relates to 1) tools for capturing and visualising data
provenance information, and 2) existing efforts to standardise ex-post
harmonisation. For encoding design, this work borrows heavily from 3)
multi-layer graph visualisations and bijective mapping techniques and
tools.

\hypertarget{data-preprocessing-documentation-and-visualisation}{%
\subsection{Data Preprocessing Documentation and
Visualisation}\label{data-preprocessing-documentation-and-visualisation}}

The documentation of data preprocessing decisions is important for the
evaluation and interpretation of analysis based on a given dataset. Data
preprocessing decisions can greatly impact downstream results within a
given project, as well as the suitability of a dataset for use outside
that project. Reflecting the importance of these decisions, prompts for
describing preprocessing steps are often included in templates for
documenting datasets. \emph{Datasets for Datasheets}
\citep{gebruDatasheetsDatasets2021} and \emph{Data Cards}
\citep{pushkarnaDataCardsPurposeful2022} both include prompts for
capturing preprocessing steps such as parsing, cleaning,
transformations, sampling and labelling. However, these frameworks aim
to capture information about all aspects of dataset production. As a
result, there is limited guidance on how to capture and present more
complex task or domain specific data provenance information.

A similarly broad approach is evident in data provenance visualisation
tools. Bors et al.~note that ``the instrumentation of provenance in
visual analytics applications often concerns only sub-tasks of
analysis'' \citep[ p.2]{borsProvenanceTaskAbstraction2019} and attribute
this to difficulty of inferring higher level tasks from lower-level data
manipulations. Many tools focus on visualising the directed analytic
graph (DAG) produced by the step-wise manipulation of a dataset.
Smallset Timelines \citep{lucchesiSmallsetTimelinesVisual2022} samples
rows from before and after selected preprocessing steps which are then
rendered as a static timeline of dataset snapshots. On the interactive
side, \emph{VisTrails} \citep{callahanVisTrailsVisualizationMeets2006}
and \emph{DQProv Explorer}
\citep{borsCapturingVisualizingProvenance2019} both provide step-wise
provenance graph views. DAG-style tools are most suited to dataset
manipulations which can be summarised as a chain of step-wise functions.
This limits their applicability to more complex transformation steps
such as cross-taxonomy transformations.

\hypertarget{task-abstraction-for-ex-post-harmonisation}{%
\subsection{Task Abstraction for Ex-Post
Harmonisation}\label{task-abstraction-for-ex-post-harmonisation}}

The process of transforming related datasets and merging them into a
single analysis dataset has various names, including ex-post (or
retrospective) data harmonisation
\citep{ehlingHarmonisingDataOfficial2003, kolczynskaCombiningMultipleSurvey2022}.
Perhaps owing to the scarcity of data relative to the breadth of
research questions, ex-post harmonisation is particularly common in the
social sciences. For instance, it is commonly used on survey data from
different countries or time periods to produce panel datasets. Panel
datasets are longitudinal datasets with repeated observations of the
same units across time. Producing panel datasets from related data often
requires both sequential transformations, such as to resolve
inconsistencies between time periods for a given country unit; as well
as concurrent transformations of multiple country units into a common
taxonomic schema for given time period.

Depending on the complexity of mappings used, and the number of data
sources, ex-post harmonisation can be both time-consuming to implement
and difficult to document. The core challenge is that the required data
transformations may not be simple one-to-one functions of data frames.
In particular, consider a two-column data table containing name-value
pairs where the name column contains a categorical index such as
occupation code, and the values encode some numeric mass associated with
a given index category. Examples of such index-value pairs include
counts of workers by occupation, value of production by industry, or
population by administrative area. Only in the simple case of one-to-one
category recoding can the data transformation step be summarised as a
function, i.e.~a renaming function which uniquely transforms the name
half of each index-value pair. In the presence of any other type of
relation, such as many-to-one or one-to-many, the data transformation
step ceases to be a simple function. Instead multiple concurrent
transformations across subsets of the index-value pairs are required.

\begin{figure}[t]

{\centering \includegraphics{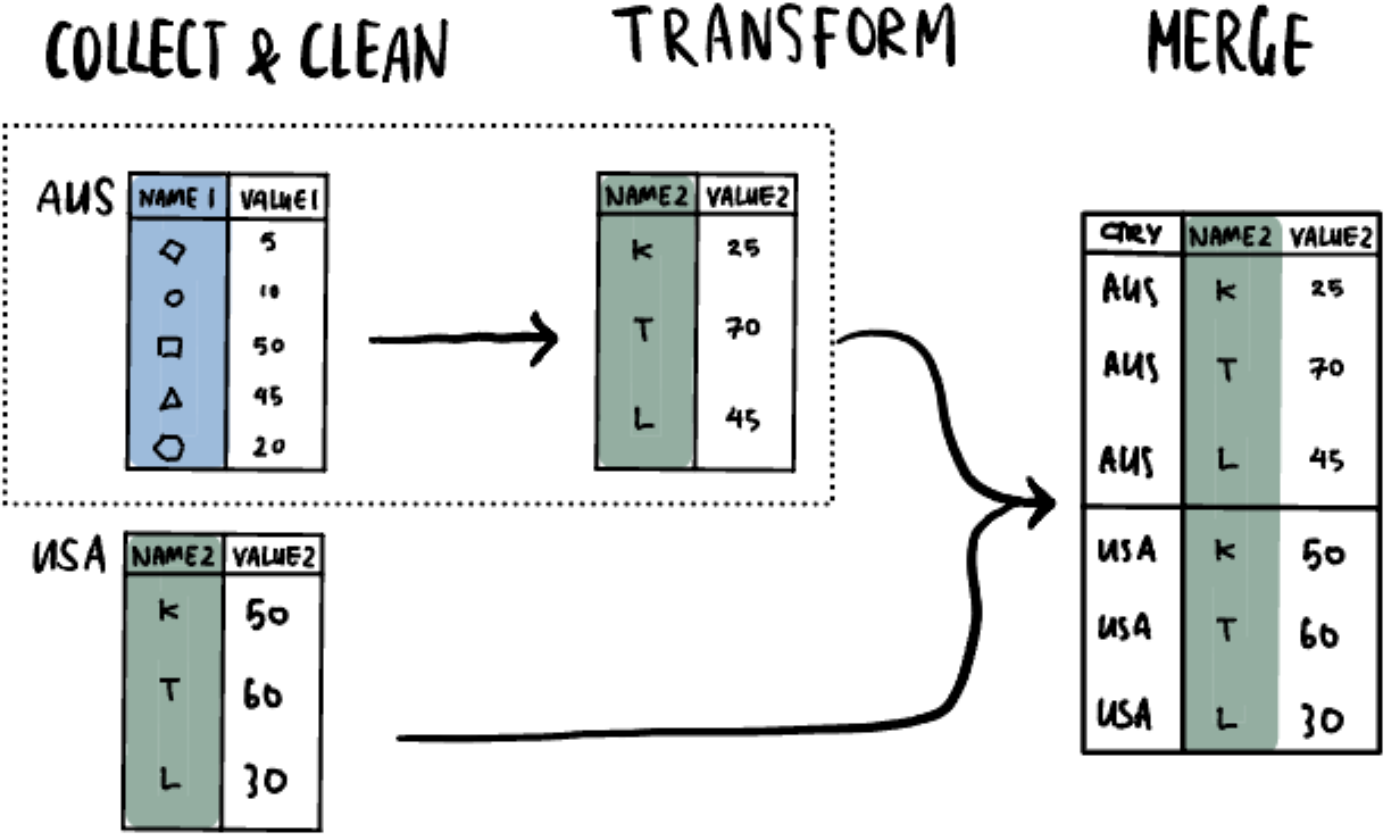}

}

\caption{\label{fig-ex-post-process}Decomposition of Ex-Post
Harmonisation Process}

\end{figure}

Figure~\ref{fig-ex-post-process} illustrates the harmonisation of
stylised data from two observational units, AUS and USA. For each
observational unit, some semantically related numeric data (e.g.~count
of workers) has been observed, but under different taxonomies (i.e.~blue
vs green occupation codes). To combine the numeric data for the two
observational units into a single dataset, a index-value mapping from
the blue to green taxonomy is selected or designed. If each blue
category can be mapped to one and only one green category, the implied
value mapping is one-to-one. However, if a single blue category is
meaningfully related to two or more green categories, the redistribution
of values must be explicitly specified.

The process of ex-post harmonisation can be decomposed into the
following steps:

\begin{enumerate}
\def\labelenumi{\arabic{enumi}.}
\tightlist
\item
  \textbf{Data Collection:} discovering and obtaining datasets
  containing harmonisable data
\item
  \textbf{Source Specific Preparation:} identifying and resolving issues
  specific to a data source and collection method (i.e.~missing values,
  reshaping, renaming variables etc.)
\item
  \textbf{Cross-Taxonomy Transformation:} transforming each source
  dataset into a common taxonomy. This stage can be further split into
  the design or selection of the harmonisation scheme and the actual
  data manipulation.
\item
  \textbf{Data Merging:} merging the transformed data into a single
  analysis-ready dataset.
\end{enumerate}

Using Bors et al.'s hierarchical provenance task abstraction framework
\citep{borsProvenanceTaskAbstraction2019}, we distinguish low-level,
highly granular data manipulation tasks performed across all steps, from
the higher level abstract task of mapping and transforming labels and
data between taxonomies. The higher level task is what we refer to as a
\emph{cross-taxonomy transformation}. We choose the infrequently used
descriptor `Cross-Taxonomy' over more common terms such as `Schema' or
`Categorical' transformations to avoid confusion with established usage.
Schema transformations in the database literature include a broader set
of tasks than addressed here, including matching at the metadata level
(i.e.~renaming columns). Cross-taxonomy transformations include but are
not limited to recoding nominal or ordinal categorical variables. We use
the term \emph{categorical recoding} to refer to the simpler task of
renaming, adding or collapsing of index labels between taxonomies. We
further use the term \emph{numeric redistribution} to refer to the
modification of associated numeric data through aggregation or
disaggregation.

A common example of a cross-taxonomy transformation from applied
economics involves converting occupation statistics from a
country-specific standard into the International Standard Classification
of Occupations (ISCO). For instance, Humlum \citep{humlum2022robot}
harmonises and integrates Danish micro-data with occupation codes from
the 1988 and 2008 versions of the Statistics Denmark's Classification of
Occupations (DISCO88 and DISCO08) to study interactions between robot
adoption and labour market dynamics. In the absence of an officially
published DISCO88 to DISCO08 correspondence, Humlum creates a custom
mapping to perform ex-post harmonisation. Humlum combines multiple
published correspondences from both the International Labour
Organisation and Statistics Denmark with relations inferred from job
code changes in the microdata.

Unfortunately, details of such mappings are often hidden away in custom
data wrangling scripts, and only described in general terms as part of
the data preparation process. Detailed notes on the DISCO88-DISCO08
correspondence created for and used to prepare the analysis data are not
included in the main paper or appendix, and can only be found in a
separate documentation note on the author's website
\citep{humlumCrosswalksISCO88ISCO082021}. Furthermore, code and data
availability standards
\citep[e.g.][]{korenmiklosDataCodeAvailability2022} only partially
support the auditing and reuse of cross-taxonomy transformations.
Although highly transparent, custom scripts predominantly capture
low-level data manipulation decisions. It can be difficult to separate
and reuse the domain specific expertise embedded in the cross-taxonomy
transformation from the idiosyncrasies of the data wrangling tool or
approach used to complete the task.

\hypertarget{crossmaps-for-cross-taxonomy-transformations}{%
\subsection{Crossmaps for Cross-Taxonomy
Transformations}\label{crossmaps-for-cross-taxonomy-transformations}}

There are some existing efforts towards standardising ex-post
harmonisation to support code comprehension and reuse. Crosswalk based
workflows are often recommended
\citep[e.g.][]{kolczynskaCombiningMultipleSurvey2022}. A crosswalk (or
concordance) table consists of at least two columns, one containing
categories in the source taxonomy, and another with the target index.
Additional information such as category descriptions, or flags for
partial (one-to-many) relations may also be included. Except for cases
where there are only one-to-one relations, crosswalks are lateral
(i.e.~one-way) mappings. Table 1 shows a simple one-to-one crosswalk
between different ISO 3166-1 country code sets with implied value
redistribution between all sets being one-to-one. Each row encodes a
pair-wise relations between codes (i.e.~categories) in each of the
country code standards (i.e.~taxonomies).

However, the crosswalk approach cannot handle cross-taxonomy
transformations where a single category is related to more than one
category in the target taxonomy, otherwise known as a one-to-many
relation. The pairwise lookup table structure of crosswalks cannot
capture additional decisions around how to distribute data between
categories. To overcome this limitation, first note that the
category-wise specification of cross-taxonomy transformations are
naturally suited to layered graph representations. A cross-taxonomy
transformation between two taxonomies can be fully described by a
directed-bipartite weighted graph, whereby the categories of the source
and target taxonomies are treated as the disjoint node sets, edges
between the source and target layers indicate which categories are
related to each other, and weights on the edges specify how numeric
values should be distributed between layers. We can extend this
representation to multiple sequential transformations, by concatenating
multiple graphs into a multi-partite graph. For concurrent
transformations of multiple datasets to a common taxonomy, we may have a
collection of graphs which all share the same target layer.

\begin{table}
\caption{\label{tbl-iso-crosswalk}Crosswalk for converting country codes}\tabularnewline
\centering
\begin{tabular}{l|l|l|l}
\hline
country & ISO2 & ISO3 & ISONumeric\\
\hline
Afghanistan & AF & AFG & 004\\
\hline
Albania & AL & ALB & 008\\
\hline
Algeria & DZ & DZA & 012\\
\hline
American Samoa & AS & ASM & 016\\
\hline
Andorra & AD & AND & 020\\
\hline
\end{tabular}
\end{table}

\begin{figure}[hb]

{\centering \includegraphics{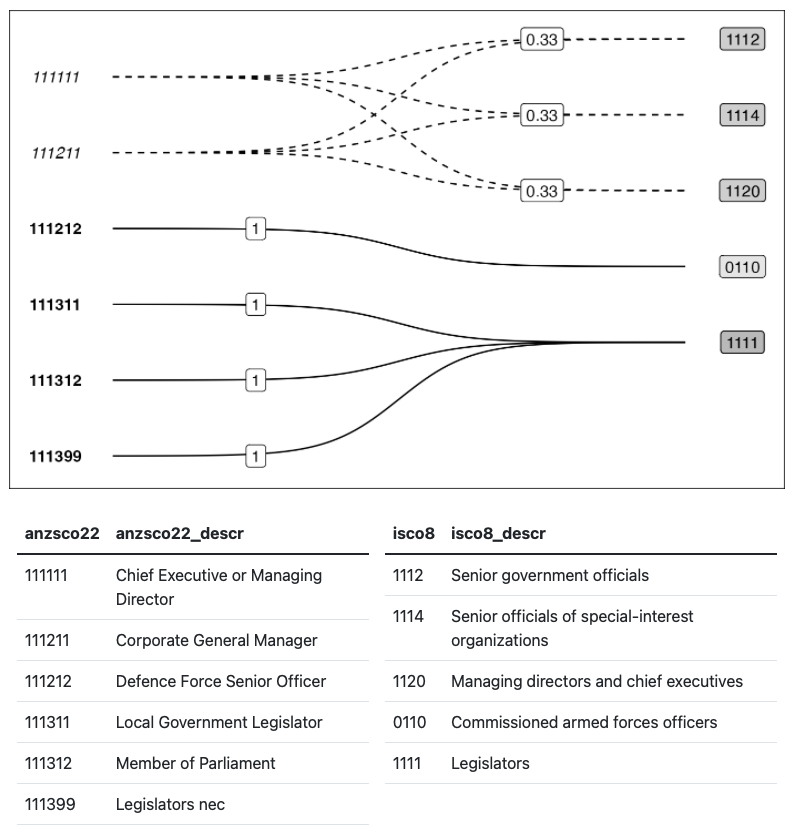}

}

\caption{\label{fig-job-crossmap}ANZSCO22 to ISCO8 Crossmap}

\end{figure}

We propose a new graph-based information structure, the \emph{crossmap},
which extends the crosswalk structure to support weighted relations
between taxonomies. Crossmaps are a class of directed, multi-partite,
weighted graphs, with properties which ensure they describe valid
recoding and redistribution from one taxonomy to another. Table 2 shows
the tabular or edge-list form of a crossmap. The crossmap structure
appends a column of weights to the source and target columns of a
standard crosswalk. Weights can take on any value between 0 and 1, with
0 implied by the absence of a link. Thus, each row represents a directed
relation between a source and target category, with unit or fractional
weights denoting what share of numeric mass associated with a source
category should be distributed to the target category.
Figure~\ref{fig-job-crossmap} illustrates a subset of a crossmap based
on a crosswalk between the 2022 update of the Australian and New Zealand
Standard Classification of Occupations (ANZSCO22) and the fourth
iteration of the International Standard Classification of Occupations
(ISCO08) published by the Australian Bureau of Statistics
\citep{ABSANZSCOAustralianNew2022}.

\vfill\null

\vfill\null

In a similar fashion to crosswalks, crossmaps serve as lookup tables for
cross-taxonomy transformations with the following data manipulation
steps:

\begin{enumerate}
\def\labelenumi{\arabic{enumi}.}
\tightlist
\item
  Rename original category names into target categories.
\item
  Multiply source values by link weight.
\item
  Summarise multiplied values by target category to obtain the
  transformed index-value variable.
\end{enumerate}

Additionally, conditions on node degree and weights dictate what type of
cross-taxonomy transformation a crossmap encodes. For example, a
crossmap where all source and target nodes have a degree of one, and all
links have unit weights, describes a one-to-one renaming transformation
(i.e.~crosswalks are a special case of crossmaps). For transformations
with a combination of one-to-one and one-to-many relations, we can
ensure numeric mass (i.e.~column totals) is preserved by verifying the
following properties:

\begin{enumerate}
\def\labelenumi{\arabic{enumi}.}
\tightlist
\item
  There is at most one link between each distinct source and target
  category
\item
  For each source category, the sum of weights attached to all outgoing
  links sums to one.
\end{enumerate}

Details of the transformation workflow and validity conditions are
omitted for brevity, as this work focuses on the visualisation of the
crossmap structure.

\begin{table}
\caption{\label{tbl-ctry-crossmap}Crossmap for recoding and distributing country statistics}\tabularnewline
\centering
\begin{tabular}{l|l|r}
\hline
from & to & weight\\
\hline
BLX & BEL & 0.5\\
\hline
BLX & LUX & 0.5\\
\hline
E.GER & DEU & 1.0\\
\hline
W.GER & DEU & 1.0\\
\hline
AUS & AUS & 1.0\\
\hline
\end{tabular}
\end{table}

\begin{figure*}[hbt]

\begin{minipage}[b]{0.20\linewidth}

{\centering 

\includegraphics{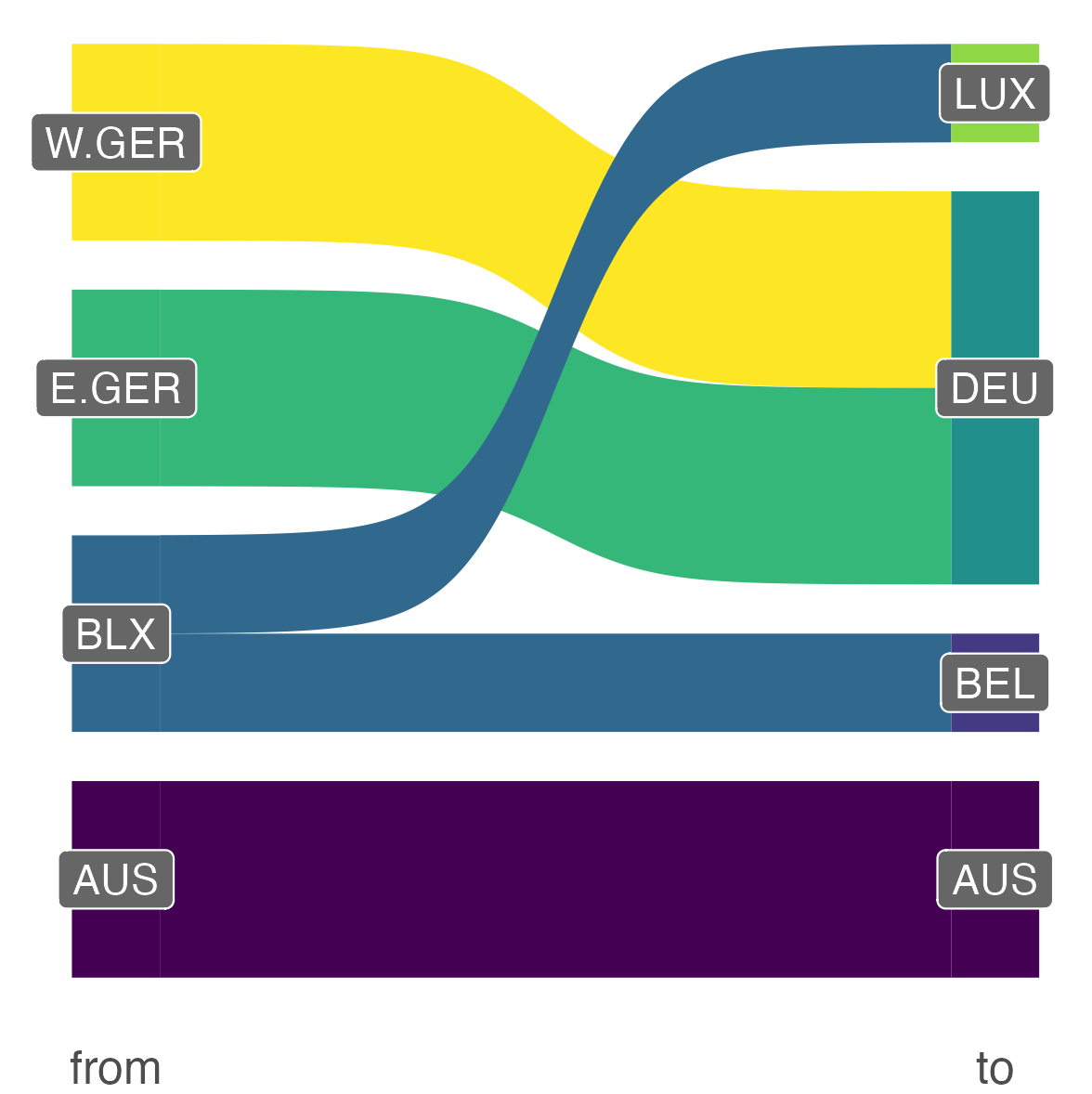}

}

\subcaption{\label{fig-ggsankey}Sankey}
\end{minipage}%
\begin{minipage}[b]{0.30\linewidth}

{\centering 

\includegraphics{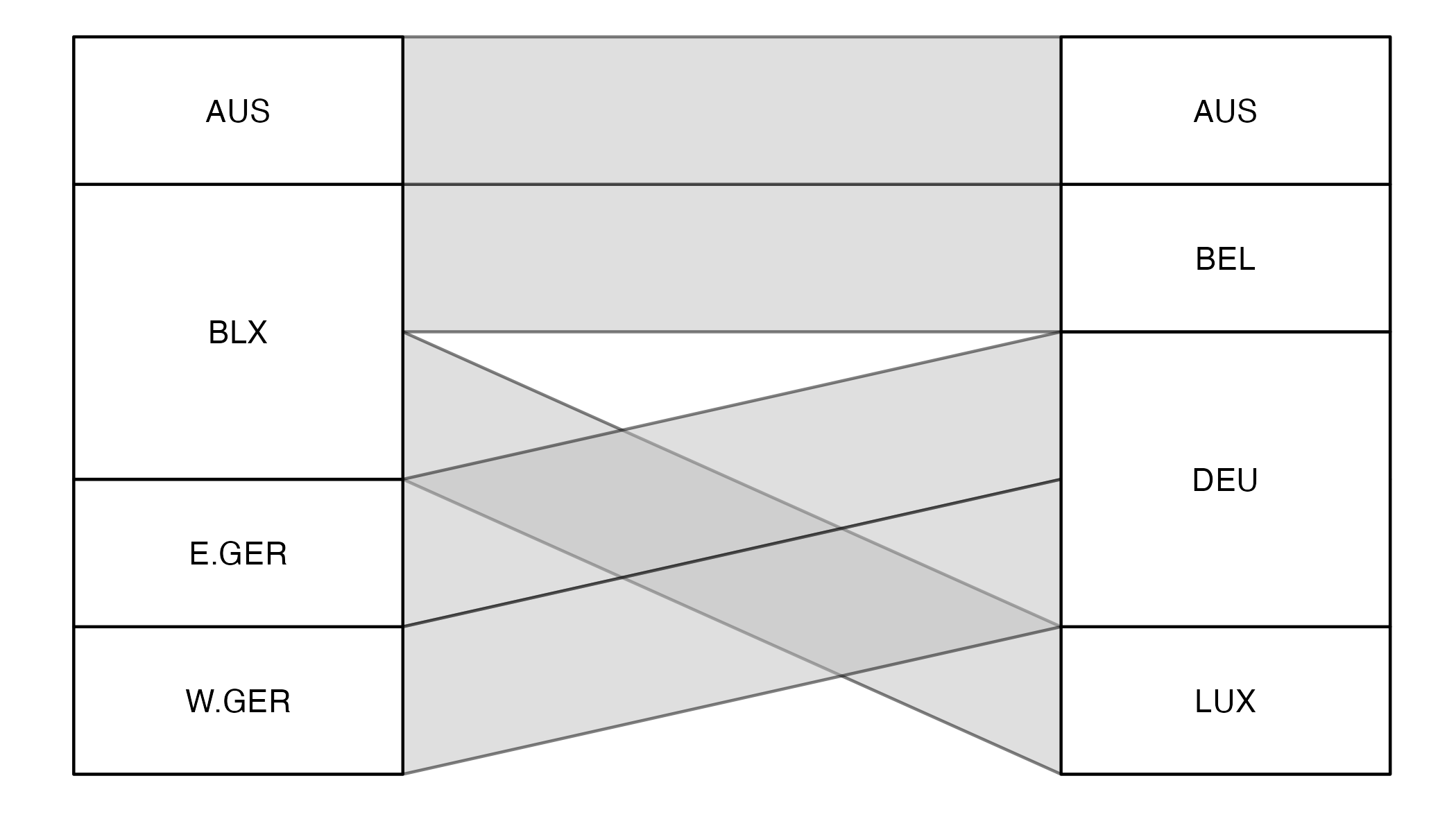}

}

\subcaption{\label{fig-ggalluvial}Alluvial}
\end{minipage}%
\begin{minipage}[b]{0.50\linewidth}

{\centering 

\includegraphics{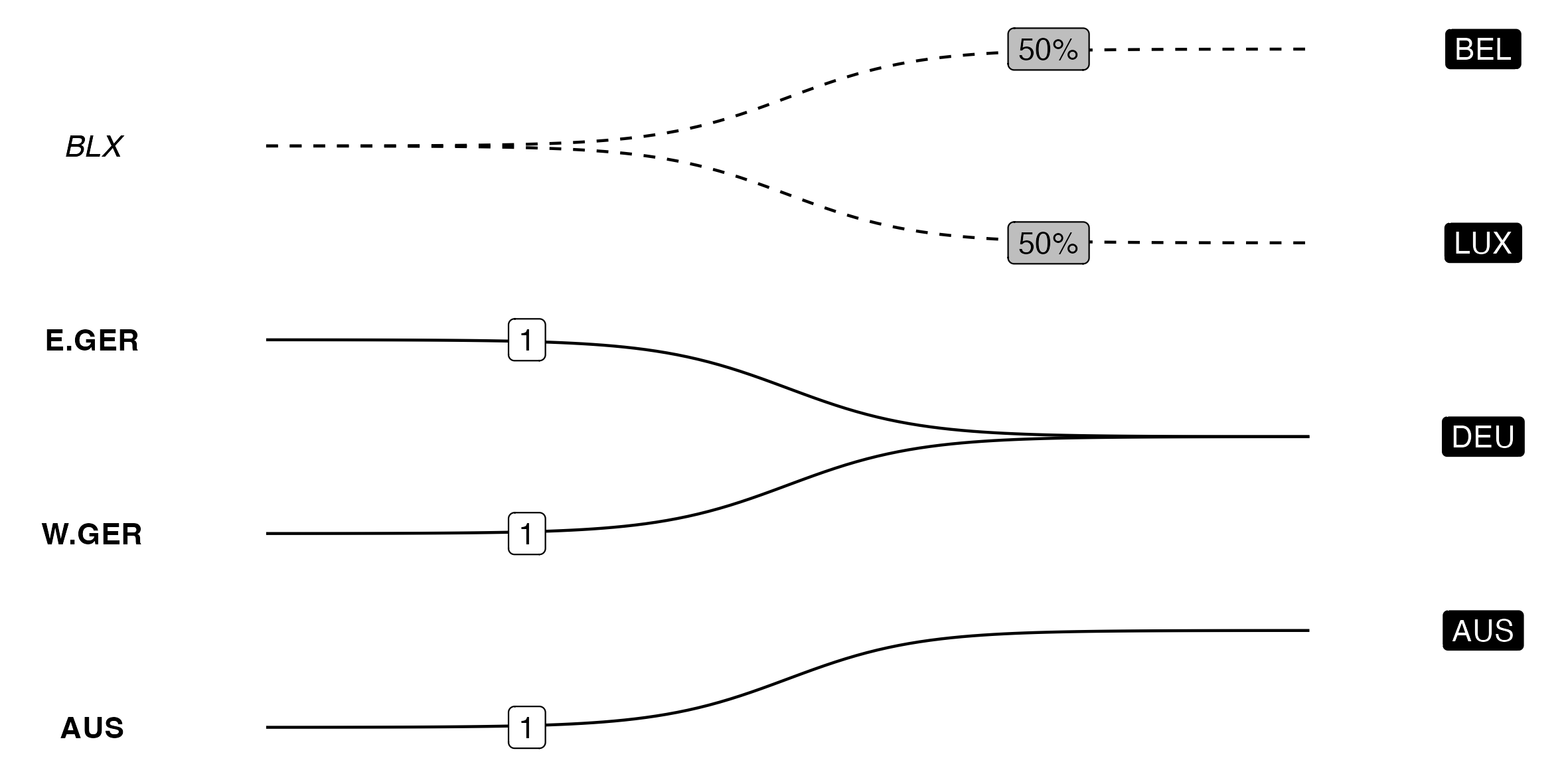}

}

\subcaption{\label{fig-ggbump}Node-Link}
\end{minipage}%

\caption{\label{fig-blx}Three representations of the crossmap in Table
2.}

\end{figure*}

\hypertarget{role-of-visualisation}{%
\subsection{Role of Visualisation}\label{role-of-visualisation}}

We focus primarily on visualisation as a tool for communicating data
provenance decisions. The target user and audience for the crossmap
structure and associated visualisations are consumers of analysis based
on ex-post harmonised data. Their primary interests are likely to be the
robustness and replicability of the final analysis, which could include
estimates of some causal coefficients or predictions of future outcomes.
This is in contrast to bijective or multi-layer visualisation tools
which support more exploratory or design focused processes such as
taxonomy alignment
\citep[e.g.][]{dangProvenanceMatrixVisualizationTool2015}, schema
matching, or sense-making
\citep[e.g.][]{sunBiSetSemanticEdge2016, fiauxBixplorerVisualAnalytics2013}.
For ex-post harmonisation, officially published correspondences are
generally preferred over novel alignments or mappings. Custom mappings
are generally only created in cases where an official correspondence
does not exist, or ambiguous one-to-many mappings are resolved with
additional redistribution weights to transform numeric data.
Furthermore, these ambiguities may only affect a small subset of
categories, and thus may not warrant more complex inferential processes
or alignment tools.

Crossmaps have much stronger constraints than generic data provenance
DAGs. As such, relative to provenance graphs, the problem of visualising
crossmaps shares more similarities with the visualisation of
genealogical graphs, as discussed by McGuffin and Balakrishnan
\citep{mcguffinInteractiveVisualizationGenealogical2005}. Similar to
family trees, crossmaps have special structural properties which can be
exploited and highlighted in visual representations. These structures
include, potentially overlapping, sub-trees corresponding to one-to-many
(split) or many-to-one (aggregate) relations; as well as sub-graphs
corresponding to many-to-many relations whereby a subset of source
categories corresponds to a subset of target categories in aggregate,
but each individual source categories correspond to multiple target
categories. Moreover, crossmaps are a class of multi-partite graph, for
which layered graph visualisation techniques such as Sankey diagrams and
Sugiyama-style graph drawings are designed for. Given the presence of
non-trivial structures, we outline two ways visualisation can support
the data provenance value of crossmaps: 1) highlighting connections with
a common underlying logic, and 2) illuminating which nodes and
connections modify the source data the most.

Firstly, visualisation can support communication and validation of the
assumptions underpinning a crossmap. Highlighting and/or grouping
connections that share a common underlying logic can help to surface and
summarise the domain expertise embedded in the mapping. By definition,
applying one-to-many splits involves more subjectivity than one-to-one
category renaming. In addition to deciding which categories map to each
other, a secondary question of how to distribute data between multiple
target categories must be answered. Visually distinguishing between
one-to-one and one-to-many links highlights this additional question,
inviting viewers to examine one-to-many links more closely. Secondly,
visualisation can also highlight when and where source data will be most
heavily manipulated by a given crossmap. For instance, the number of
incoming for a given target category can be used as a measure of how
synthetic the final value associated with that category is.
Visualisation can thus highlight the decisions that are most likely to
affect the robustness of downstream analyses.

\hypertarget{visualising-crossmaps}{%
\section{Visualising Crossmaps}\label{visualising-crossmaps}}

\hypertarget{existing-graphical-representations}{%
\subsection{Existing Graphical
Representations}\label{existing-graphical-representations}}

Since the crossmap structure extends upon crosswalk tables, it is useful
to examine existing methods for visualising crosswalks. Sankey diagrams
and alluvial diagrams both have multi-layer characteristics that make
them common choices for visualising crosswalks
\citep[e.g.][]{wuAnalysisCrosswalksResearch2023}. However, variable node
sizes and link widths can actually clutter the visualisation of
crosswalks. Consider the Sankey and Alluvial diagrams in
Figure~\ref{fig-ggsankey} and Figure~\ref{fig-ggalluvial}, produced
using the \texttt{ggplot2} \citep{wickhamGgplot2ElegantGraphics2016}
extensions \texttt{ggsankey} \citep{ggsankey-package} and
\texttt{ggalluvial} \citep{ggalluvial-package} respectively. They both
represent a crossmap for harmonising national accounts data (e.g.~GDP)
across two time periods. However, compared to the node-link style
diagram in Figure~\ref{fig-ggbump}, produced using \texttt{ggbump}
\citep{ggbump-package}, it is less obvious that the correspondence
involves an equal (50-50) split relation from \texttt{BLX} to
\texttt{\{BEL,\ LUX\}}. Furthermore, although the variable node heights
in Figure~\ref{fig-ggalluvial} and Figure~\ref{fig-ggsankey} partially
reflect different relation types, the node-link layout more clearly
distinguishes the one-to-many split.

\hypertarget{proposed-visualisation}{%
\subsection{Proposed Visualisation}\label{proposed-visualisation}}

Our proposed node-link style visualisation for crossmaps is illustrated
in Figure~\ref{fig-job-crossmap}. Starting from the source category
nodes on the left, we use text style to distinguish categories which
will be redistributed (italics) from categories with values that will be
unmodified (bold) in the first two data manipulation steps. We reiterate
this distinction using line style, with dashed links for splits and
solid links for one-to-one outgoing relations. Next, we use text labels
to show link weights, noting that unit weights could be omitted to
further distinguish splits. The stylised examples shown in
Figure~\ref{fig-job-crossmap} and Figure~\ref{fig-blx} both apply equal
weight splits for one-to-many relations. However, it is both more common
and more realistic to have custom unequal weighting schemes based on
some reference information. For example, population or GDP statistics
can be used to estimate or inform disaggregation shares. Finally, we
style the target nodes using colour or opacity to distinguish target
categories based on the number of incoming contributions. The
combination of incoming link style and target node darkness helps to
convey how synthetic a given target category is.

In addition to node-link properties, Figure~\ref{fig-job-crossmap}
illustrates a number of layout and scalability considerations. Firstly,
note that we place source nodes with one-to-many outgoing relations
above one-to-one source categories. Again, this directs attention
towards redistribution weights and any associated assumptions. However,
if we wanted to highlight the composition of the transformed data, we
could arrange the layout based on the incoming degree of the target
nodes. The optimal layout depends both on structural characteristics of
the crossmap such as what types of relations are present, as well as the
purpose of the visualisation. Second, issues of readability and label
crowding arise not only as the size of the crossmap grows (i.e.~number
of nodes and links), but also with features like category name length
and the presence of many-to-many relations. Overlapping weight labels
can be seen between \texttt{\{111111,\ 111211\}} and
\texttt{\{1112,\ 1114,\ 1120\}}, while a companion table is used to
avoid printing long and variable length node names.

Figure~\ref{fig-job-crossmap} was produced as part of a vignette in
v0.0.1 of the \texttt{xmap} package \citep{xmap-pkg} using the
\texttt{ggplot2} extension \texttt{ggbump}. The \texttt{xmap} package is
a prototype implementation of the crossmap structure and related
functions in R. v0.0.2 of \texttt{xmap} includes functions to generate
similar graphics from valid crossmaps using the \texttt{ggraph}
\citep{pedersenGgraphImplementationGrammar2022} extension. For
multi-layer graphs, \texttt{ggraph} only offers Sugiyama's
\citep{sugiyamaMethodsVisualUnderstanding1981} heuristic layout
algorithm as implemented by \texttt{igraph}
\citep{csardiIgraphSoftwarePackage2006}. Unfortunately, this means
additional modifications are required to sort nodes and links by the
criteria discussed above. However, we plan to improve the visualisation
functions in future versions of \texttt{xmap}.

\hypertarget{discussion}{%
\section{Discussion}\label{discussion}}

\hypertarget{threats-to-validity}{%
\subsection{Threats to Validity}\label{threats-to-validity}}

Munzner \citep{munznerNestedModelVisualization2009} emphasises the need
to address threats to validity at each stage of visualisation design. To
this end, we note that visualisations are not commonly used to
communicate information about ex-post harmonisation. This is perhaps due
in part to challenges in characterising the process of ex-post
harmonisation and designing useful abstractions. As such, the
abstractions and visualisations described in this paper are yet to be
validated with the target audience of social science researchers.
However, the crossmap structure and associated prototype implementation,
\texttt{xmap}, have been informed by informal discussions with applied
social science researchers who conduct ex-post harmonisation and the
author's own experience producing harmonised datasets for economic
research. This work also illustrates challenges and opportunities
associated with capturing and communicating mid-level provenance tasks.

\hypertarget{future-extensions-and-applications}{%
\subsection{Future Extensions and
Applications}\label{future-extensions-and-applications}}

This paper focuses predominantly on static visualisations of single-step
transformations between single layer taxonomies. However, the task
abstraction and visualisation design can also be extended to multi-step
visualisations, interactive tools and other types of data.

For sequential multi-step transformations, the meaning and visual
encoding of source and target node properties would need to be extended
for intermediate layers. Furthermore, additional layers give rise to new
layout design challenges. Rather than simply minimising edge crossings,
multi-step crossmap layouts should also incorporate link type or
weights, as well as any relevant grouping structures such as category
hierarchies and/or useful sub-graphs. These layout challenges share some
similarities with optimising layouts for Sankey diagrams as discussed by
Zarate et al. \citep{zarateOptimalSankeyDiagrams2018}, and further work
could look at adapting their Integer Linear Programming model to
crossmaps. Another type of multi-step transformation involves the
concurrent transformation of multiple datasets with different source
taxonomies into a single target taxonomy. The visualisation challenges
include displaying multiple crossmaps in a coherent manner and how to
exploit the fact that each crossmap shares the same target node layer.

Currently, the \texttt{xmap} package only supports input of crossmaps
via edge lists, thus limiting the accessibility of the provenance
structure to individuals who can code. Interactive tools for visualising
and modifying crossmaps could make the crossmap structure more
accessible to domain experts. Node label readability, crowding and
scalability issues inherent in larger multi-step crossmaps can also be
addressed using interactive features such as tooltips and collapsible
sub-graphs Further work on identifying meaningful sub-structures in
crossmaps could be also used to design small multiple visualisations or
filter interactive views. For example, one-to-one relations could be
collapsed by bundling links and grouping nodes using bicluster
visualisation techniques \citep[e.g][]{sunBiSetSemanticEdge2016}.

Although the crossmap structure was designed to capture cross-taxonomy
mapping decisions for tabular data in ex-post harmonisation, the
underlying structure can be used to capture mapping decisions in other
similar preprocessing procedures. For example, crossmaps could be used
to encode redistribution of population statistics across two overlapping
sets of geographic units such as postal codes and census tracts. Postal
codes and census tracts form the two disjoint layers in our crossmap,
and distribution decisions can again be captured by link weights. Given
the spatial dimension of the data, it would also be useful to explore
map based visual representations in addition to the node-link graphics
discussed here.

\hypertarget{conclusion}{%
\section{Conclusion}\label{conclusion}}

We presented a novel task abstraction for categorical recoding and
redistribution transformations in the context of ex-post data
harmonisation, which we termed \emph{cross-taxonomy transformations}. We
described a graph structure, the \emph{crossmap}, for capturing
provenance information about this task. Crossmaps support the separation
of domain expertise from lower-level data manipulation actions in
cross-taxonomy transformations. Node, link and structural properties of
crossmaps reflect key characteristics of these transformations such as
the presence of one-to-many splits or many-to-one aggregations. We
visualised these properties in a two-layer node-link style graphic and
demonstrated how different visual channels can be used to highlight
particular transformation characteristics and identify significant
source and target categories. We concluded with a discussion of validity
threats and future directions.

\acknowledgments{We thank Laura Puzzello for her ongoing support and
funding of earlier iterations of this work. Many thanks also to Rob
Hyndman, Sarah Goodwin, Emi Tanaka, Patrick Li, Simon Angus and my other
colleagues at Monash EBS and Monash SoDa Labs for their helpful
guidance, feedback and suggestions. We also thank the anonymous
reviewers for their constructive comments. The author is supported in
part by top-up scholarships from Monash Data Futures Institute and the
Statistical Society of Australia.}

\bibliographystyle{abbrv-doi-hyperref}

\bibliography{references.bib}

\appendix 

\end{document}